# Some information on acoustic topological insulator

*Partha Goswami and Udai Prakash Tyagi*
*D.B.College, University of Delhi, Kalkaji, New Delhi-110019, India*

**Abstract**   In this exceedingly short review article, we have provided some information on acoustic topological insulator (ATI) for pedagogical purpose. Since, intrinsically acoustic systems do not have Kramers doublets due to spin-0 status, artificially acoustic spin-1/2 states could be engineered (ref.5) as reported in refs. 6-26 maintaining time reversal symmetry. The high point of this article is an explanation of emergent Dirac physics in ATIs.

**Main Text**
The topological insulators (TIs) and their higher order variants [1-4] are materials that are insulating in the bulk but have topologically protected, conducting surface states. This requires time-reversal symmetry to be preserved, apart from the momentum-spin coupling. Their energy dispersion relation exhibits a Dirac cone shape. This scenario ensures no backscattering of electrons. The acoustic topological insulators (ATI)[5-26] are similar to conventional electronic topological insulators. They can be described using a hierarchy of topological invariants known as the Dirac hierarchy (DH). Finding an ATI material, however, is quite burdensome. One possible approach, to realize a 2D ATI with protected chiral edge states and a Dirac cone (DC)-shaped surface dispersion relation, is to use the geometry of conventional TIs to create artificial ones. The latter may be composite materials made up of a lattice of coupled resonators forming a 2D structure. The frequencies of these resonators could be made sensitive to the direction of the incident wave. This leads to the symmetry breaking and a topologically non-trivial band structure that can support surface states with DCs. A lattice configuration of a series of Helmholtz resonators, which produces an acoustic equivalent of the Haldane model (for electronic TIs), also falls in the category of acoustic TIs. Furthermore, upon using anisotropic materials/ geometries, such as graphene-like materials or layered composites, one introduces a structural asymmetry in the material. This also helps to realize two-dimensional surface DCs as these systems break the symmetry of the acoustic medium. For example, a honeycomb lattice structured phononic crystal. The crystal lattice is made up of two materials with one of them being denser than the other. A symmetry breaking is ensured by the resultant structural anisotropy yielding DC in certain directions of the Brillouin zone. Another method is to use surface acoustic waves on a thin, piezoelectric film, where the film generates an electric field perpendicular to the surface that breaks the mirror symmetry. Yet another approach is to create a periodic pattern of subwavelength holes on the surface of a material. These can lead to an effective medium that hosts surface acoustic waves with a linear dispersion relation and a Dirac-like point with the caveat that the size and spacing of the subwavelength holes are properly tuned.

The recent report [18] demonstrates an acoustic 3D honeycomb lattice that features a Dirac hierarchy comprising an eightfold bulk DC, a 2D fourfold surface state DC, and a 1D twofold hinge state DC all using the same 3D base lattice.  The 3D lattice appears as a first-order TI with 2D topological surface states, a second-order TI exhibiting 1D hinge states, and a third-order TI of zero-dimensional mid-gap corner states due to the lifting of the Dirac degeneracy in each hierarchy. They analytically discuss the topological origin of the surface, hinge, and corner states in this communication. The states correspond to out-of-plane and in-plane winding numbers. The investigations on higher order(acoustic) topological insulator(HOTI) could be found in refs. [19-26]. The study of Xinglong Yu et al [18] offers new approach to control sound and vibration for acoustic steering and guiding. It also takes care of on-chip ultrasonic energy concentration, filtering, etc.. The design procedures followed by them is based on the bottom-up approach of band-structure engineering. Their study definitely expands the fundamental research scope of Dirac physics and multidimensional robust wave manipulation. We shall get back to details of this work later.

The one-dimensional hinge state DC, in particular, leads to hinge transport which is a topologically protected sound wave transport that occurs in ATI [5-26]. The hinge transport is referred to as the ability of excitations to propagate along edges. The edges are formed due to the intersection of three-dimensional topological insulators with their surrounding environments or vacuum. The lowest level invariant in DH relates to the existence of gapless surface states, robust against disorder and other perturbations, protected by the time-reversal symmetry (TRS). As is well-known [1-4], any backscattering between the counter-propagating modes on the surface is forbidden due to TRS. At higher levels of DH in ATI [5-26], multiple pairs of energy bands that cross each other at isolated points in momentum space. These isolated points are referred to as multiple Dirac points (MDP). They cannot be removed or split into smaller points without breaking the symmetry of the system. This means that the MDPs are protected by topology. In fact, the presence of MDPs in the energy spectrum of a three dimensional ATI with DH induces the appearance of hinge states that allow for the propagation of excitations along the edge. To achieve one-dimensional hinge DC in acoustic crystal, one will need to follow the steps given below: First, one needs to select a phononic crystal that can exhibit an one dimensional hinge DC with a suitable bandgap. Then analyse the band structure using finite element simulations or other tools that can accurately model acoustic wave propagation. The selection process requires using techniques such as acoustic microscopy or optical measurement methods. In order to maximize the DC size/ bandwidth and optimization of the crystal performance, one needs to tune the parameters of the crystal. The crystal may be fabricated using acoustic metamaterials or other suitable materials. In order to do the characterization of the crystal, one can perform experiments to investigate its acoustic signal transmission, reflection, and absorption properties. Furthermore, one should be able to adjust the lattice constant, the size of the scatterers, the material properties of the crystal, etc. depending on the application. It must be emphasized that careful control of the effect of defects and imperfections in the crystal is very important as even small variations in the crystal structure or material properties can significantly affect the behaviour of the Dirac cones and other acoustic phenomena.

In general, topological insulators (TIs) are characterized by a bulk gap and in-gap boundary states. The conventional $d$-dimensional TIs are featured with $(d-1)$ dimensional boundary states [1,2,24]. For instance, the one-dimensional TIs, such as that described by the Su-Schrieffer-Heeger (SSH) model, host zero-dimensional end states; the two-dimensional TIs host one-dimensional edge states, such as the quantum spin Hall insulators. Recently, however, it was reported that the higher-order TIs exhibit an extended bulk-boundary correspondence: a $d$- dimensional -$n^{th}$ order TIs have $(d-n)$ dimensional boundary states [27,28,29]. For example, a three-dimensional third order TI (or a two-dimensional second-order TI) possesses the zero-dimensional corner states. Some of these compelling physical effects have been extended to artificial band gap materials, viz. photonic crystals [30,31] and phononic crystals [32-35].

In order to investigate the ATI system theoretically, one needs to write the corresponding Hamiltonian $H_{3D}$ in the form $H_{3D} = \sum_k c^{\dagger}_k H_{3D}(k) c_k$, where $c_k$ is the vector of acoustic equivalent of annihilation operators and $H_{3D}(k)$ is the (acoustic equivalent) Bloch matrix. This will indicate the basis chosen (to write $H_{3D}(k)$) of the sublattices in the n-th unit cell. The effective tight binding model of an acoustic crystal is a theoretical model that describes the behaviour of the crystal's vibrations in terms of the interaction between its constituent atoms or molecules. In this model, the crystal is treated as a lattice of discrete points, or "sites," each of which represents an atomic or molecular unit. The strength of the interaction between neighbouring sites is characterized by a set of parameters, such as the hopping energy, which describes the energy required for a phonon to move from one site to another. These parameters are typically determined by fitting the model to experimental data or more detailed calculations, and can be used to predict the phonon dispersion relations and density of states for the crystal. The effective tight binding model is useful for studying the properties of acoustic crystals because it provides a quantitative description of their vibrations and can be used to predict a variety of phenomena, such as phonon localization, band gaps, and resonance modes. It has been applied to a wide range of materials, including one- and two-dimensional structures, as well as more complex three-dimensional crystal lattices. The model is particularly useful for predicting the properties of acoustic metamaterials, which are engineered structures designed to have unusual or exotic acoustic properties, such as negative refraction or acoustic cloaking. The effective tight binding model can also be extended to include additional interactions between the lattice sites, such as anharmonic terms or long-range interactions. These extensions can improve the accuracy of the model and allow for the prediction of more complex phenomena, such as phonon-phonon interactions and nonlinear effects. Overall, the effective tight binding model of an acoustic crystal is a powerful tool for understanding the behaviour of these materials and designing new acoustic metamaterials with tailored properties. However, it is important to note that the model has its limitations and simplifications. For instance, it assumes that the lattice vibrations can be described as discrete phonons, whereas in reality, the vibrations are continuous and can involve a range of frequencies. Additionally, the model may not fully capture the effects of disorder or defects in the crystal lattice. Therefore, the effective tight binding model should be used in combination with other experimental and theoretical techniques to fully characterize the behaviour of acoustic crystals. Other techniques that could be used to complement this model include spectroscopy, scanning probe microscopy, and density functional theory.

The tight-binding Su-Schrieffer-Heeger model (SSH model is one of the simplest models of topology and the resulting edge states i.e. the bulk-boundary correspondence in the presence of the chiral symmetry) takes the form, for a molecular chain, as bipartite Hamiltonian $H^{SSH} = -\sum_n(t_1 |n,A\rangle \langle n,B| + t_2|n+1,A\rangle \langle n,B| + h.c)$, where $n$ labels the unit cell and $A$ and $B$ are the different sublattices with hopping parameters $t_1$ and $t_2$ (say, $t_2 \sim 0.75 t_1$) between them. For example, Polyacetylene consists of a long chain of carbon atoms with alternating single and double bonds between them, each with one hydrogen atom. The double bonds can have either *cis* or *trans* geometry. The *cis* form of the polymer is thermodynamically less stable than the *trans* isomer. Thus, two configurations for polyacetylene lead to different hopping parameters between carbon atoms. We can label the chain with two sublattices A and B, and we have a unit cell that includes two atoms. We write the corresponding extended $k$-space Hamiltonian matrix out as $H_k^{SSH} = -(t_1 + t_2 \cos(k))\sigma_x - t_2 \sin(k) \sigma_y + E(k)\sigma_0 + M\sigma_z$ where $\sigma_j'$s are Pauli matrices and E(k)/M is the additional term breaking the chiral symmetry/ time reversal symmetry. All these facts are well known [36]. An effective Hamiltonian to analyse the nature of the acoustic crystal is, in fact, a $k$-space 12× 12 Hamiltonian matrix [18] which could written as the one

($H_{3D}(k)$) involving four objects $H_{3D}(k) = \begin{bmatrix} \boldsymbol{\varepsilon_0}(k) & \boldsymbol{M}(k) \\ \boldsymbol{M^*}(k) & \boldsymbol{\varepsilon_0}(k) \end{bmatrix}$ where the objects $\boldsymbol{\varepsilon_0}(k)$ and $\boldsymbol{M}(k)$ are 6 × 6 matrices involving transfer integrals. This is a generalization of the extended SSH model $H_k^{SSH}$ for the problem of ATI investigated by Xinglong Yu et al.[18]. Apart from the translation symmetry, which has been used to write $H_k^{SSH}$, the other symmetries in SSH model $H_k^{SSH}$ ($E(k) = 0$, $M = 0$) are Time reversal symmetry, Chiral symmetry, Particle-hole symmetry, and Inversion symmetry. The study of Xinglong Yu et al.[18] leads to a comprehensive understanding of the hinge states along the three independent directions in the three dimensional-HOTIs.

In conclusion, there is a pertinent issue that whereas the realization of the HOTIs are only possible in the artificial classical structures defined under the Hermitian condition, such systems involve intrinsic loss which makes the corresponding Hamiltonian non-Hermitian. Kawabata et al.[37] have reported higher-order non-Hermitian skin effect. The detailed report of an experimental demonstration of non-Hermiticity-induced HOTI was provided by He Gao et al.[38]. This possibly brings to light a new horizon in the study of non-Hermitian physics and the design of non-Hermitian systems.